\documentstyle[sprocl]{article}

\input{epsf}

\begin{document}

\title{
{\raggedleft  IITAP-99-013 \\
hep-ph/9911228 \\}
{ \bf The Next-to-Leading  BFKL Pomeron \\
with Optimal Renormalization}
\footnote{ \textit{ presented by V. T. K. 
at the VIIIth International Blois Workshop,
June 28 - July 2, 1999, Institute for High Energy Physics, 
Protvino, Russia, to appear in the Proceedings
}}}

\author{ \bf Victor~T.~Kim${}^{\ddagger \&}$, Lev~N.~Lipatov${}^{\ddagger}$ \\
{ \rm and} \\
Grigorii~B.~Pivovarov${}^{\S \&}$ }
 
\address{${}^\ddagger$ : St.Petersburg Nuclear Physics Institute,  188350 Gatchina,
Russia 
\newline
${}^\S$ : Institute for Nuclear Research, 117312 Moscow, Russia 
\newline
${}^\&$ : International Institute of Theoretical and Applied Physics, \\
Iowa State University, Ames,
IA 50011, USA}

\maketitle
\abstracts{
The  next-to-leading order (NLO) corrections to  the BFKL equation 
in the BLM optimal scale setting are briefly discussed.
A striking feature of the BLM approach 
is rather weak $Q^2$-dependence of the Pomeron intercept,
which might indicate an approximate conformal symmetry of the equation. 
An application of the NLO BFKL resummation  for the virtual gamma-gamma 
total cross section shows a good agreement
with recent L3 data at the CERN LEP2.}

The Balitsky-Fadin-Kuraev-Lipatov (BFKL) \cite{BFKL,BL78} 
resummation of energy 
logarithms is anticipated to be an important tool for exploring 
the high-energy limit of QCD.  
Namely, the highest eigenvalue, $\omega^{max}$, of the BFKL 
equation \cite{BFKL} is 
related to the intercept of the Pomeron which in turn governs 
the high-energy asymptotics of the cross sections: $\sigma \sim 
s^{\alpha_{I \negthinspace P}-1} = s^{\omega^{max}}$. 
The BFKL Pomeron intercept in the leading order (LO) 
turns out to be rather large: 
$\alpha_{I \negthinspace P} - 1 =\omega_{LO}^{max} = 
12 \, \ln2 \, ( \alpha_S/\pi )  \simeq 0.55 $ for 
$\alpha_S=0.2$; hence, it is very important to know the next-to-leading order
(NLO) corrections.
In addition, the LO BFKL calculations have restricted phenomenological 
applications because, {\it e.g.}, the running  of the QCD coupling 
constant $\alpha_S$ is not included, and the kinematic range of 
validity of LO BFKL is not known.

Recently the  NLO corrections to the BFKL resummation
of energy logarithms were calculated; see Refs. \cite{FL,CC98} and references
therein. The NLO corrections \cite{FL,CC98} to the highest eigenvalue 
of the BFKL equation turn out to be negative and even larger
than the LO contribution for $\alpha_S > 0.16$. 

It should be stressed that the NLO calculations,
as any finite-order perturbative results, contain 
both  renormalization scheme and renormalization scale ambiguities.
The NLO BFKL calculations \cite{FL,CC98} were 
performed by employing the modified minimal subtraction scheme 
($\overline{\mbox{MS}}$) to regulate the ultraviolet 
divergences with arbitrary scale setting.

In the recent work \cite{BFKLP} it was found that 
the renormalization scheme dependence of the NLO BFKL resummation of 
energy logarithms  \cite{FL,CC98} is not strong, {\it i.e.},
value of the NLO BFKL term is practically the same
in the known renormalization schemes.
To resolve the renormalization scale ambiguity due to the large 
NLO BFKL term \cite{FL,CC98} the  
Brodsky-Lepage-Mackenzie (BLM) optimal scale setting \cite{BLM} has been utilized. 
The BLM optimal scale setting effectively resums the conformal-violating
 $\beta_0$-terms into the running coupling in all orders 
of the perturbation theory.

It was shown \cite{BFKLP} that the reliability of QCD predictions 
for the effective intercept of the BFKL Pomeron at NLO evaluated at the  
BLM scale setting within the non-Abelian
physical schemes, such as the momentum space 
subtraction (MOM) scheme 
or the $\Upsilon$-scheme based on  $\Upsilon \rightarrow ggg$ decay, 
is significantly improved compared to the $\overline{\mbox{MS}}$-scheme result 
\cite{Ross,Blu98}.

One of the striking features of the analysis \cite{BFKLP} is that the NLO value for 
the intercept of the BFKL Pomeron, improved by the BLM procedure, has a 
very weak dependence on the gluon virtuality $Q^2$: 
$\alpha_{I \negthinspace P} - 1 =\omega_{NLO}^{max} =   \simeq 0.13 - 0.18 $
at $Q^2 = 1 - 100$ GeV$^2$.
It arises as a result of fine-tuned compensation between the LO and NLO contributions.
The minor $Q^2$-dependence obtained leads
to approximate conformal invariance.

As a phenomenological application of the NLO BFKL
improved by BLM procedure one can consider the gamma-gamma scattering \cite{BFKLP99}.
This process is attractive because
it is theoretically more under control than the hadron-hadron
and lepton-hadron collisions, where non-perturbative hadronic structure
functions are involved. In addition, in the gamma-gamma scattering
the unitarization (screening) corrections
due to multiple Pomeron exchange would be less important than in
hadron collisions.

The gamma-gamma cross sections with the BFKL resummation in the LO  
was considered in \cite{BL78,Brodsky97,Bartels96}.
In the NLO BFKL case one should obtain a formula analogous to
LO BFKL \cite{BFKLP99}. 
While exact NLO impact factor of gamma is not known yet, 
one can use the LO impact factor of  \cite{BL78,Brodsky97} assuming
that the main energy-dependent  NLO corrections come from the NLO
BFKL subprocess rather than from the impact factors \cite{BFKLP99}.

\begin{figure}[htb]
\begin{center}
\leavevmode
{\epsfxsize=8.5cm\epsfysize=8.5cm\epsfbox{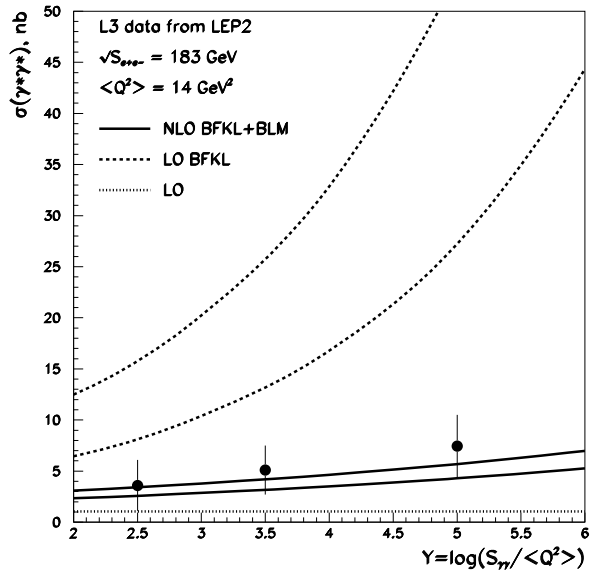}}
\end{center}
\caption[*]{
Virtual gamma-gamma total cross section
by the NLO BFKL Pomeron within BLM approach vs L3 Collaboration data
at energy 183 GeV of $e^{+}e^{-}$ collisions.
Solid curves correspond to NLO BFKL in BLM; dashed: LO BFKL
and dotted: LO contribution. Two different curves are for 
two different choices of the Regge scale: $s_0= Q^2/2$, $s_0=2 Q^2$}
\label{fig:4}
\end{figure}

In Fig. \ref{fig:4} the comparison of BFKL predictions for LO and NLO
BFKL improved by the BLM procedure with L3 data 
\cite{L3} from CERN LEP2 is shown. The different curves reflect uncertainty with the
choice of the Regge scale parameter which defines 
the beginning of the asymptotic regime.
For present calculations two variants have been choosen $s_0=Q^2 / 2$
and $s_0 = 2 Q^2$, where $Q^2$ is virtuality of photons. 
One can see from Fig. \ref{fig:4} 
that the agreement of the NLO BFKL improved by the BLM procedure 
is reasonably good at LEP2 energy 
$\sqrt{s_{e^+e^-}}=$ 183 GeV.
One can notice also that sensitivity of the NLO BFKL results
to the Regge parameter $s_0$ is much smaller than in 
the case of the LO BFKL.

 It was shown in Refs. \cite{Kaidalov86,Cudell99}
that the unitarization corrections in hadron collisions
can lead to higher value of the (bare) Pomeron intercept than
the effective intercept value. Since the hadronic data fit yields
about 1.1 for the effective intercept value \cite{Cudell99}, 
then the bare Pomeron intercept value should be above this value. 
Therefore, assuming small unitarization corrections in the gamma-gamma
scattering at large $Q^2$ one can accomodate the NLO BFKL Pomeron
intercept value 1.13 - 1.18 \cite{BFKLP} in the BLM 
optimal scale setting along with larger unitarization corrections 
in hadronic scattering \cite{Kaidalov86}, where they can lead to 
a smaller effective Pomeron intercept value about 1.1 for 
hadronic collisions. 

Another possible application of the BFKL approach
can be the collision energy dependence of the inclusive 
single jet production \cite{KP98}.
 
There have been a number of recent papers which analyze the NLO BFKL
predictions 
in terms of  rapidity correlations 
\cite{Lipatov98},
angle-ordering \cite{CCFM},  double transverse momentum 
logarithms \cite{Andersson96,Salam98,Ciafaloni99}, 
an additional $\log(1/x)$ enhancement \cite{Ball99} 
and  BLM scale setting 
for deep inelastic structure functions  \cite{Thorne99}.
Obviously, a lot of work should be done to clarify the proper expansion parameter 
for BFKL regime and, also the relation between
those papers and the result of the present BLM approach.
To confirm the result of \cite{FL,CC98} the independent NLO calculations 
(see  \cite{Duca99} and references therein) 
for BFKL resummation  are desirable. 

To conclude, we have shown that the NLO corrections to the BFKL 
equation for the QCD Pomeron become controllable
and meaningful provided one uses physical renormalization scales and
schemes relevant to non-Abelian gauge theory.  BLM optimal scale setting
automatically sets the appropriate physical renormalization scale by
absorbing the non-conformal $\beta$-dependent coefficients.   The
strong renormalization scheme and scale dependence of the NLO 
corrections to BFKL resummation then largely disappears.  
A striking feature of the NLO BFKL Pomeron intercept in 
the BLM approach is its very weak $Q^2$-dependence, which provides
approximate conformal invariance. The NLO BFKL application
to the total gamma-gamma cross section shows a good agreement
with the L3 Collaboration data at CERN LEP2 energies.

VTK thanks the Organizing Committee of the VIIIth Blois Workshop at
Institute for  High Energy Physics, Protvino  for their warm hospitality.
This work was supported in part by INTAS, Grant No. INTAS-97-31696.

\end{document}